%% file: jem-euso-skeleton-bibtex.tex
\documentclass[a4paper,11pt]{article}
\usepackage{pos}
\usepackage{ulem}

\usepackage{natbib}
\title{   Overview   of the JEM-EUSO program for the study of ultra-high-energy cosmic-rays from space}
 \ShortTitle{  Overview of the JEM-EUSO program}

\author*[a]{Marco Casolino}
\author[b]{Etienne Parizot}

\affiliation[a]{INFN and University of Rome Tor Vergata, \\ Italy}

\affiliation[b]{  Universit\'e de Paris, CNRS, AstroParticule et Cosmologie, F-75013 Paris, France\\
 Institut Universitaire de France (IUF), France\\}

\onbehalf{for the JEM-EUSO Collaboration\\[-1mm]{\normalsize \normalfont (a complete list of authors can be found at the end of the proceedings)}}
\emailAdd{casolino@roma2.infn.it}
\emailAdd{parizot@apc.in2p3.fr}

\abstract{Ultra High Energy Cosmic Rays (UHECRs) offer a unique chance to study the universe at 
energies  inaccessible by man-made accelerators. Observations by ground based observatories  have clarified several
characteristics of these particles, but  their origin, nature, and
acceleration mechanisms are still unclear,   mostly due to their extremely low flux. Space-based observatories  have the potential for an
increase in statistics, up to several orders of magnitude, and would be able to cover the whole sky,
allowing for a direct comparison of spectra and direction of arrival, but the detector design  poses several formidable  technical challenges.  

The JEM-EUSO program has been addressing this problem, trying to open the road of space-based UHECR observations. 
Several missions have already been completed (on the ground:
EUSO-TA; with stratospheric ballons: EUSO-Balloon, EUSO-SPB1 and EUSO-SPB2; in space: TUS\cite{Klimov2017},  MINI-EUSO). Others are under study  (K-EUSO) or proposed for the next decade (POEMMA)\cite{Olinto_2021}. In this work we report on the status of the JEM-EUSO program and the technology developed so far.  
}

\ConferenceLogo{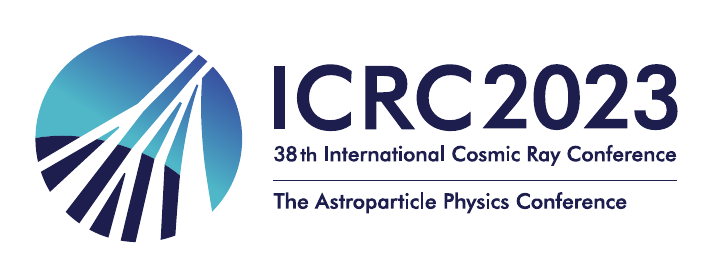}

\FullConference{%
38th International Cosmic Ray Conference (ICRC2023)\\
  26 July - 3 August, 2023\\
  Nagoya, Japan}

\begin{document}
\maketitle
\section{Introduction}

The extremely low flux of  Ultra-High-Energy Cosmic Rays (UHECR), { which are also subject to largely unknown magnetic deflections} is still preventing us from pinpointing the  location of the sources of these particles and the physics processes involved in their   acceleration. The limited statistics of the events is also preventing us from determining  their exact nuclear composition, and does not yet allow astronomy with charged cosmic rays. A space mission, devoted to the detection of the fluorescence and Cherenkov light emitted by the atmospheric showers generated by these particles, would be complementary in nature to the ground based observatories and have the potential advantage of huge sensing areas and uniform sampling of the   entire sky and thus would be valuable in overcoming several of these challenges.  
In addition to detection in the fluorescence range, detector development has progressed in the Cherenkov detection capability, enabling the search for high-energy neutrino showers. Also in this case, a space mission offers opportunities to study high-altitude, ``atmosphere-skimming'' showers, providing insights into shower physics.

In addition to the aforementioned science goals, a UHECR space mission will have various additional objectives, covering  planetary science, fundamental physics, atmospheric physics, and Earth observation. These would also include the study of meteors, transient luminous events (ELVES), ionospheric phenomena, bioluminescence, and the search for nuclearites and strange quark matter.

\section{The JEM-EUSO collaboration}

\begin{figure}[ht]
\centering
\includegraphics[width=0.48\textwidth]{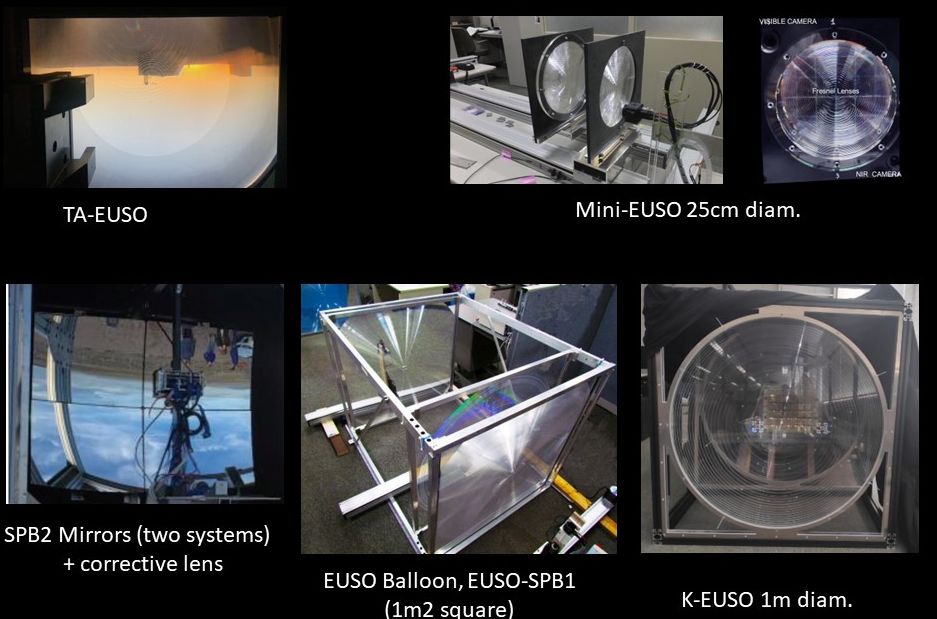}
\includegraphics[width=0.48\textwidth]{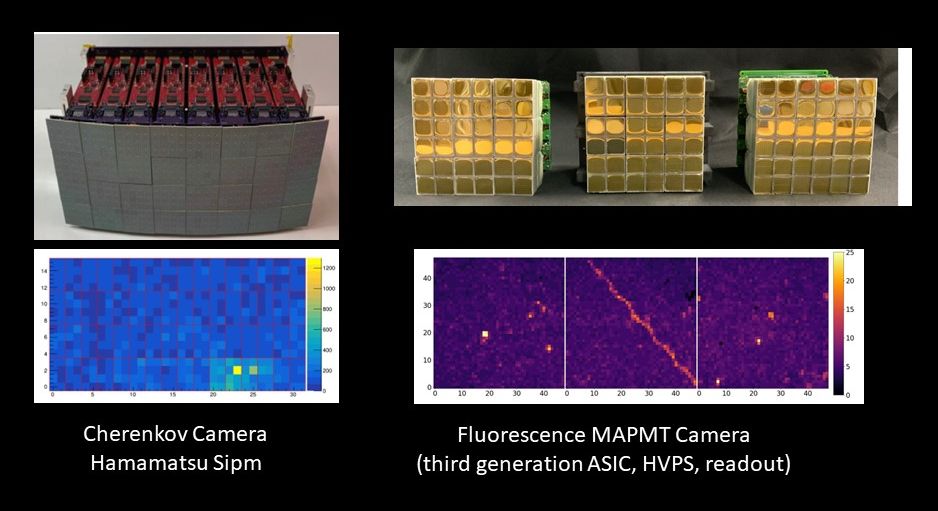}
\caption{Left: Optics developed in the framework of the JEM-EUSO collaboration. Both refractive (Fresnel-bases) and Reflective (mirror) optical systems of various sizes and focal lenghts have been developed. Right: The focal surfaces of the  Cherenkov detector and the Fluorescence detector employed in the  EUSO-SPB2 superpressure balloon payload. The Cherenkov detector camera is based on SiPM\cite{Gazda:2023cj}, whereas the Fluorescence camera\cite{Eser:2023Dw}\cite{Filippatos:2023P4}   employs Multi-Anode-Photomultipliers. }
\label{optics}       
\end{figure}

The realization of a space-based telescope capable of detecting these signals from a distance of several hundred kilometers (in the best-case low Earth orbit) has required the construction and development of various optical (Fig.~\ref{optics}) and electronics (same figure) technologies that have evolved over the years.

\begin{figure}[ht]
\centering
\includegraphics[width=0.7\textwidth]{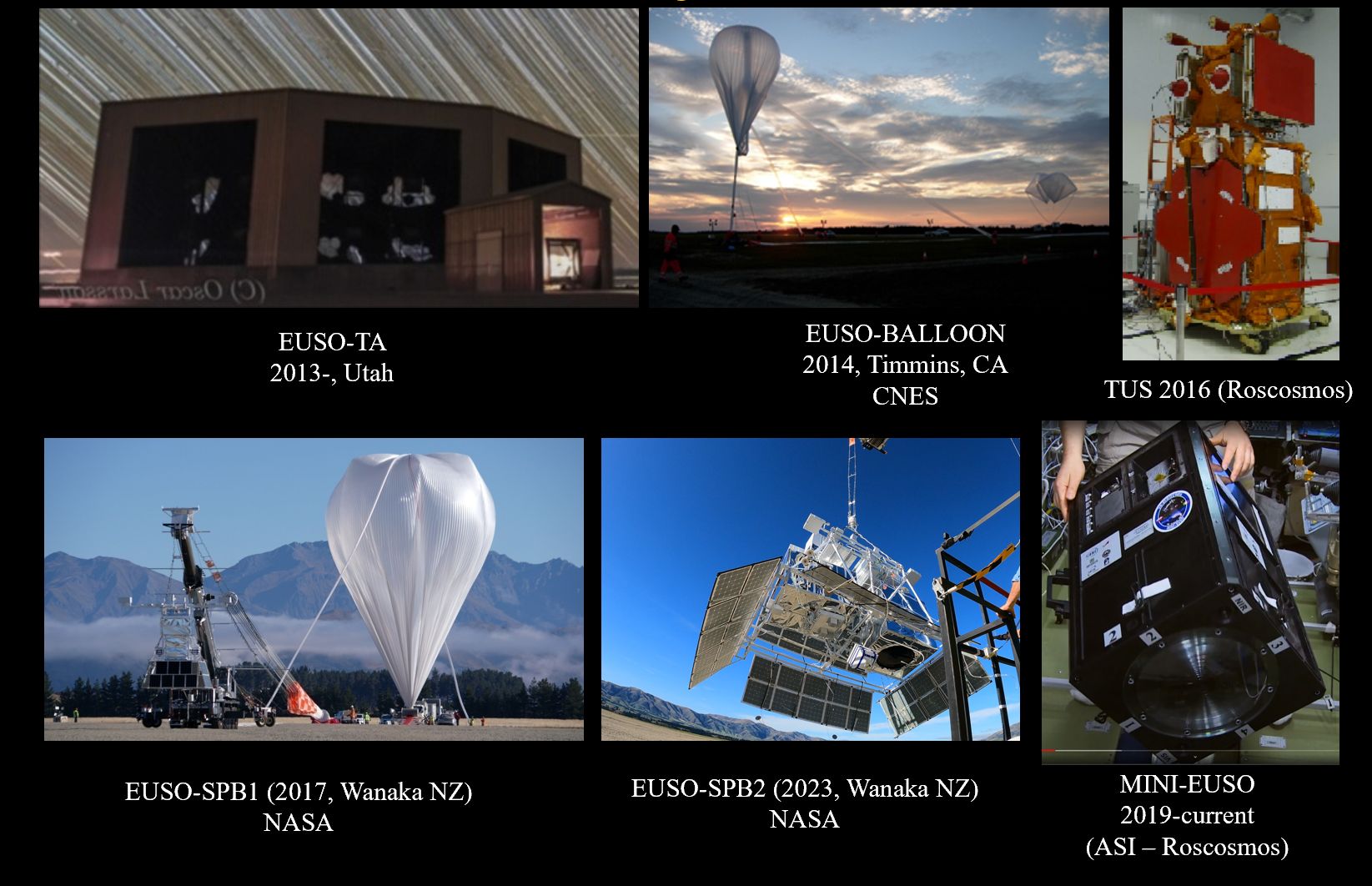}
\caption{The various projects realized by the JEM-EUSO collaboration. From left to right, top to bottom:  EUSO-TA, Ground detector installed in 2013 at the
TA site;  EUSO-BALLOON, First balloon flight from Canada, August 2014;  Russian satellite borne TUS, 2016; EUSO-SPB, Long duration
NASA flight, New Zealand, 2017;  EUSO-SPB2: 
NASA flight, New Zealand, 2023, Mini-EUSO: Earth observation telescope  installed inside
the International Space Station (ISS) in 2018.}
\label{projects}        
\end{figure}

The JEM-EUSO collaboration consists of   physicists and engineers from 12 member countries (Algeria,
Czech Republic, France, Germany, Italy, Japan, Mexico,
Poland, Russia, Slovakia, Sweden, USA) and 4 associated
countries (Republic of Korea, Romania, Spain, Switzerland). Its main goal is the development of an  Extreme Universe Space Observatory aiming at advancing the understanding of UHECRs through the observation of EAS from space. To reach this goal a number of various missions (See Figure \ref{projects}) aiming at a progressive development of the instrumentation and detection technique in view of a full JEM-EUSO mission have been, or are being, developed. In this work we briefly review the main experiments realized and the perspectives for future missions.

\section{EUSO-TA (2014-)}

\begin{figure}[ht]
\centering
\includegraphics[width=0.7\textwidth]{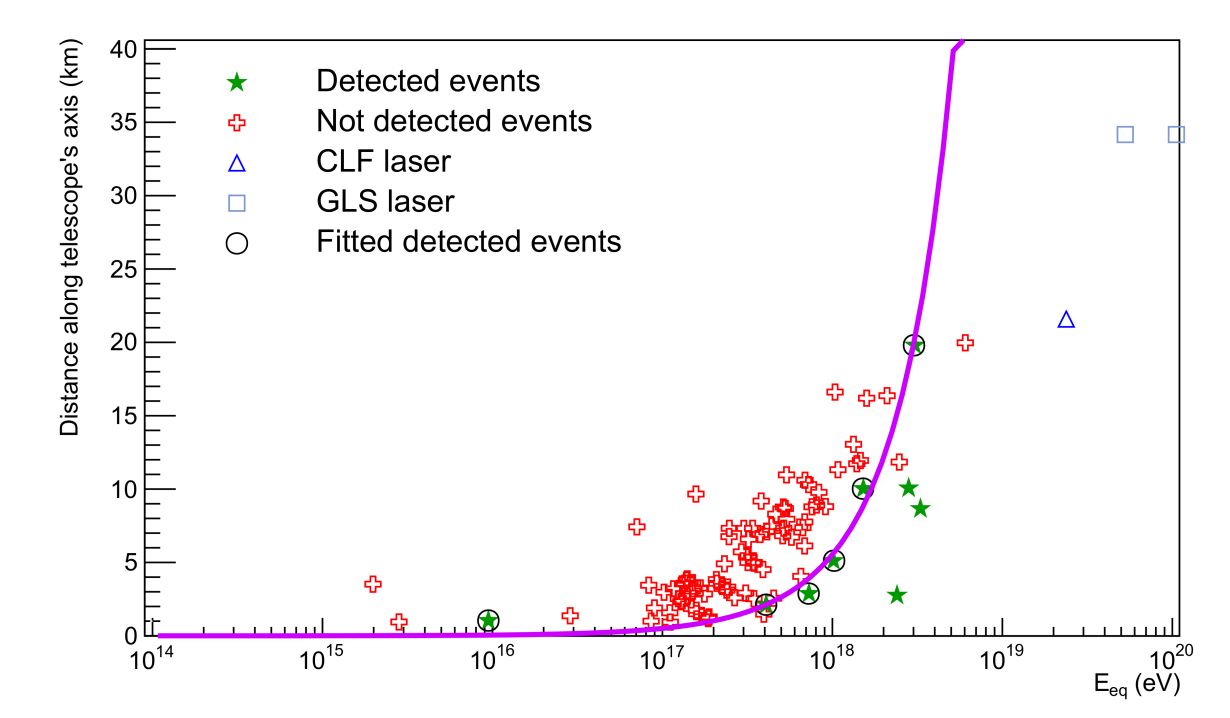}
 \caption{Distance vs Energy $E_{eq}$ of  UHECR acquired with the  EUSO-TA telescope (triggered by TA). Energy and distance come from the TA data. The purple line shows a fit of the “upper” detected events
with a second order polynomial\cite{Bisconti:2023UC}, showing the energy-distance detection capability of the EUSO-TA telescope.}
\label{eusota}       
\end{figure}

EUSO-TA is a cosmic ray detector  observing during nighttime the fluorescence light emitted
along the path of extensive air showers in the atmosphere. It is installed at the Telescope Array (TA) site in Utah,
USA, in front of the TA fluorescence detector station at Black Rock Mesa. It serves as a ground-based pathfinder
experiment for future space-based missions. EUSO-TA has an optical system with two Fresnel lenses and a
focal surface with 6 $\times $ 6 multi-anode photomultiplier tubes with 64 channels each, for a total of 2304 channels.
The overall field of view is $\simeq 10.6^{\circ} \times 10.6^{\circ} $ 
 This detector technology allows the detection of cosmic ray events
with high spatial resolution, each pixel having a field of view of about  $\simeq 0.2^{\circ} \times 0.2^{\circ} $ 
and a temporal resolution of 2.5 $\mu$s. Figure \ref{eusota} shows the energy-distance relationship for the events (triggered by TA) detected by EUSO-TA. In   2022 the system was upgraded replacing the focal surface and the acquisition system, implementing one similar to that of Mini-EUSO\cite{Plebaniak:20236g}. For further details on the calibration of the telescope see \cite{Plebaniak:20235e}, for its performance and cosmic ray detection capabilities, see \cite{Bisconti:2023UC}.

\section{Stratospheric balloon flights}
Stratospheric  balloon flights offer a unique possibility to develop the EUSO detectors and test the observation
principle in similar conditions to those encountered in space. The residual atmosphere presents an harsher environment to the high-voltage power supply system of the photomultiplier tubes (PMTs),
since a discharge is more likely than in vacuum. Also, the limited telemetry from the balloon (since
it is necessary to assume that the payload is not recovered) requires more stringent constraints to
the trigger than in the case of a space-based detector (this has been improved in EUSO-SPB2 with the use of starlink satellites). In the latter case, more bandwidth is generally
available and — in the case of the ISS — data can be physically sent to the ground on hard disks.

\subsection{EUSO-Balloon (2014)}
EUSO-Balloon, a   CNES (French Space Agency) flight , was a one-night mission with key innovations like Fresnel optics, dedicated ASIC (SPACIROC first generation) for front-end electronics, and efficient data processing. It successfully operated at 3 mbar and measured the Earth's UV emissivity at night, revealing the UV-IR luminosity anticorrelation. Laser light tracks from a helicopter flying under the balloon and Xenon flashers on the ground were also detected.

\subsection{EUSO-SPB1 (2017)}
Following EUSO-Balloon's success, the JEM-EUSO collaboration developed EUSO-SPB, a NASA mission with upgraded instrumentation and a longer flight. EUSO-SPB1 utilized a Super Pressure Balloon, enabling up to 100 days of flight time. It aimed to detect cosmic-ray showers using a focal surface with improvements, including SPACIROC-3 ASIC, more compact ECs, and enhanced optics performance. The instrument used solar panels for power supply during the extended flight and had an autonomous trigger for events.

In April 2017, EUSO-SPB1 was launched successfully from the NASA balloon facility in Wanaka, New Zealand. Although the instrument operated nominally, the flight was shortened to 12 days due to a balloon leak, preventing the detection of an actual cosmic ray shower and recovery of the instrument. Despite this, 25.1 hours of data were downloaded, providing valuable insights into the instrument's in-flight performance.

The data revealed the instrument's photometric stability within $\pm 5\%$.  It also allowed the measurement of Earth's UV emissivity, which serves as a background for the detection of extensive air showers (EAS), over various backgrounds such as land, ocean, and different types of clouds at various altitudes.

Furthermore, the data helped estimate the effective exposure as a function of energy and balloon altitude, enabling the determination of the expected event rate in different energy bins. On average, around 0.7 events were anticipated during the entire flight, which decreased to 0.4 events when considering clouds. Thus - as expected -  no events were detected.

 \subsection{EUSO-SPB2 (2022)}
 On May 13th, 2023, the Extreme Universe Space Observatory on a Super Pressure Balloon II (EUSO-SPB2) was launched from Wanaka, New Zealand. It aimed to search for very high-energy neutrinos (E>PeV) and ultra-high-energy cosmic rays (UHECRs, E>EeV) using Cherenkov radiation and ultraviolet fluorescence, respectively. The mission involved two independent optical telescopes, with the Fluorescence Telescope (FT) having 108 multi-anode photomultiplier tubes (3 PDMs) placed at the focus of a one-meter entrance diameter Schmidt  telescope\cite{Eser:2023Dw} and the Cherenkov telescope (CT) employing a Silicon Photomultiplier camera. An Infrared camera (IR) was also installed for cloud monitoring.  Prior to integration and launch, the two instruments have been field-tested in Telescope Array site, Utah\cite{Kungel:2023C6}.
 
The instrument underwent rigorous testing in both laboratory and field settings which have proven the correct functioning of all systems

Regrettably, due to a leak in the balloon the flight lasted only about 32 hours, having it sink in the Pacific Ocean. 

Despite this abbreviated flight, all instruments (FT, CT, and IR)  performed as expected, validating their functionality   anticipated from extensive simulations, laboratory, and field tests.

A substantial amount of data, approximately 56GB,  consisting of over 120,000 FT events and more than 32,000 bifocal CT triggers was downloaded in this time. Currently,  the collaboration is in the process of analyzing this data. So far, no cosmic ray candidate has been identified in the FT triggers, which aligns with the expected low rate of less than one cosmic ray event during the data collection time of less than 10 hours.

The CT data set includes several triggers from below the limb, providing valuable insights into potential neutrino observations for future missions. During approximately 45 minutes of observation above the limb, the experiment detected several Cherenkov signals from   near horizontal EAS caused by cosmic rays. This achievement not only validates the developed trigger scheme but also demonstrates the feasibility of the detection technique itself, which was another main mission objective.

\section{Mini-EUSO (2019-current)}

\begin{figure}[ht]
\centering
\includegraphics[width=0.5\textwidth]{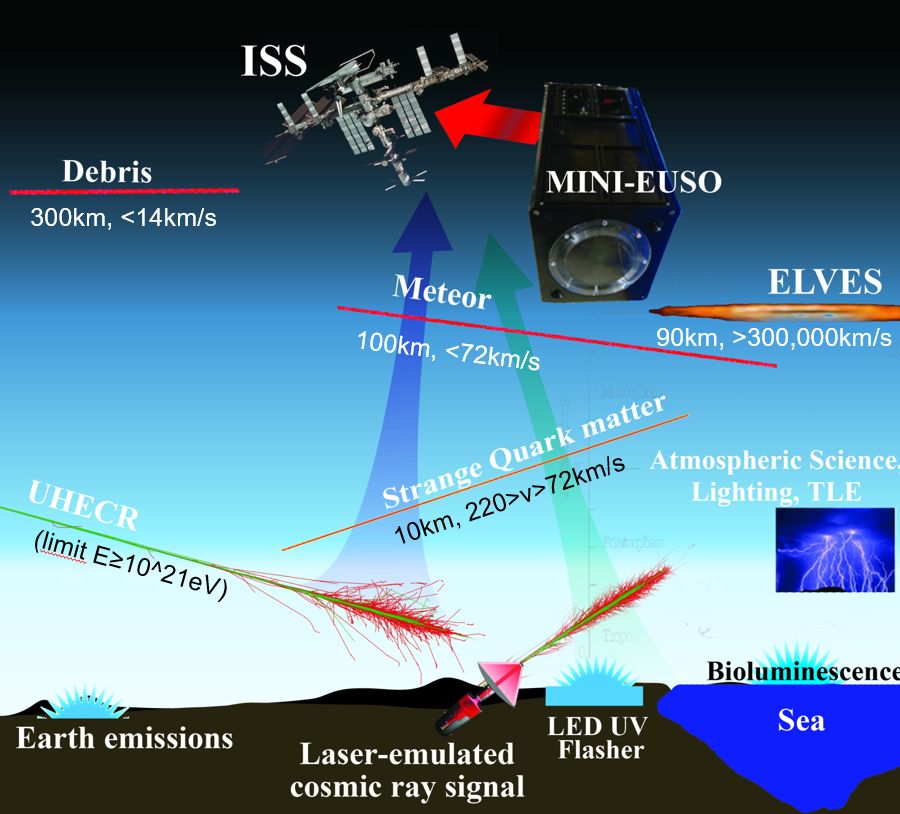}
\includegraphics[width=0.35\textwidth]{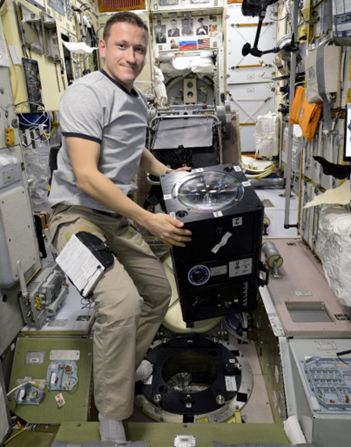}
 \caption{Left: Scientific objectives of the Mini-EUSO experiment. The detector is capable of addressing phenomena with greatly
varying intensity and duration, from the slow terrestrial emissions to the apparently superluminal ELVEs. Right: Cosmonaut Sergei Kud-Svertchkov
 installing Mini-EUSO inside the ISS on the UV-transparent window of the Zvezda module. The
round porthole on the bottom of the picture looks nadir.}
\label{science}       
\end{figure}

\begin{figure}[ht]
\centering
\includegraphics[width=0.5\textwidth]{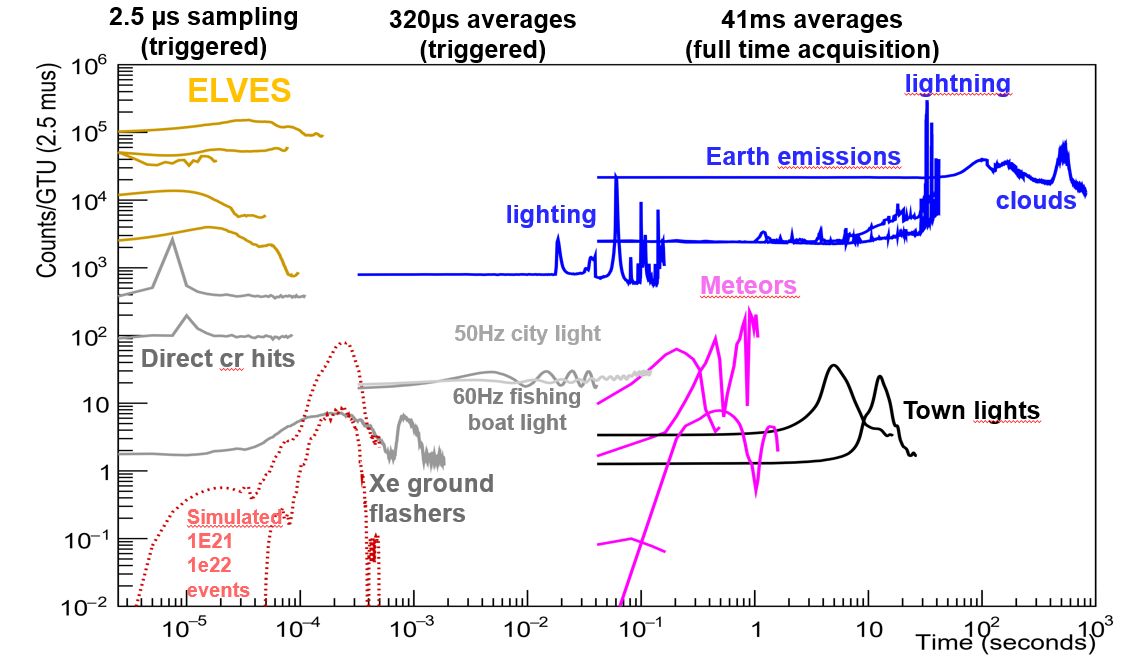}
\includegraphics[width=0.4\textwidth]{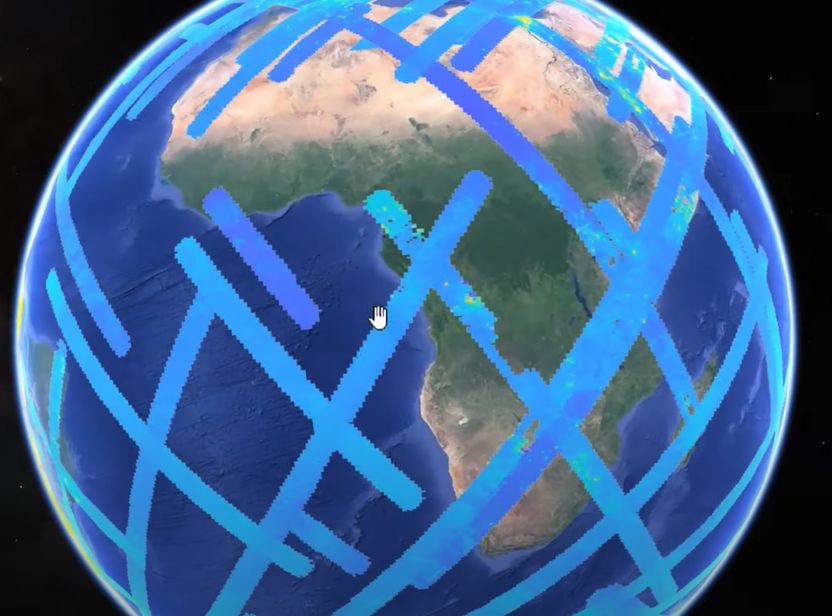}
\caption{Left: Temporal profile of various signals observed by Mini-EUSO. All curves refer to experimental data with the exception of the simulated data of UHECR at $10^{21}$   and  $10^{22}$ eV. Right: Mini-EUSO data mapped onto the surface of the Earth.}
\label{profile}       
\end{figure}

Mini-EUSO (Multiwavelength Imaging New Instrument for the Extreme Universe Space Observatory) is a telescope observing the Earth   from the International Space Station since 2019 \cite{minieuso}. The instrument  employs a  Fresnel-lens optical system and a focal surface composed of 36 Multi-Anode Photomultiplier tubes, 64 channels each, for a total of 2304 channels with single photon counting sensitivity. 
Mini-EUSO   observes the night-time Earth in the near UV range (predominantly between 290 - 430~nm) with a  projected spatial resolution  on the ground of about 6.3~km (full field of view equal to 44$^{\circ}$) and a temporal resolution of 2.5~$\mu$s, observing our planet  through a  nadir-facing UV-transparent window located  in the Russian Zvezda module.
The detector can thus  acquire triggered  transient phenomena with a sampling rate of  2.5 $\mu$s and  320~$\mu$s, as well as perform continuous acquisition at 40.96~ms scale (Fig. \ref{profile}). 
Among the  scientific objectives (Fig. \ref{science}) addressed by the mission is the mapping of night-time UV Earth \cite{CASOLINO2023113336}, the observation of meteors, the search for interstellar meteors  and amongst them, the  search for strange quark matter (SQM), appearing as anomalous (long track and fast-moving) meteor events (at the 40.96 ms time frame).  Ground-based flasher have bee also used to perform end-to-end calibration of the instrument\cite{Miyamoto:202312}. In \cite{Bertaina:2023ok} we use Mini-EUSO data to 
estimate the exposure of future space-borne UHECR detectors and in \cite{Battisti:2023kN}  to show that atmospheric phenomena cannot be mistaken for UHECR, even though their time scale is comparable \cite{CASOLINO2023113336}.

Machine learning methods have been implemented to search for meteors\cite{Bertaina:2023Tu} and ELVES\cite{Piotrowski:2023pN} with Mini-EUSO data.

\section{POEMMA-Balloon with Radio (PBR) / EUSO-SPB3 (2026) }
Although the SPB2 flight was cut short, the very good instrument performance during operations and the experience acquired allow us to  plan for a  new  balloon
mission called  POEMMA-Balloon with Radio (PBR),  also known as  EUSO-SPB3. 

The upcoming mission will improve the detectors developed in  EUSO-SPB2 with significant modifications. The main change involves merging both telescopes into a single one with a dual  focal surface, housing both the fluorescence   camera and the Cherenkov  surface camera, with tilting capacity from nadir to horizontal. The light collection area will be expanded by a factor of 1.2 to 1.5. Apart from the optical telescope, the PBR design will incorporate a radio detector for capturing   horizontal/upward-going EAS (Extensive Air Showers) produced by cosmic rays and neutrinos. This combination aims to explore the complementarity between optical Cherenkov measurements and radio detection of these events, and serve also as a pathfinder to the POEMMA Probe mission. Although the project is still in the early design phase, it is scheduled for launch from Wanaka in Spring 2026\cite{Olinto:2023ZQ}.

\section{Acknowledgements}
The authors would like to acknowledge the support by NASA award 11-APRA-0058, 16-APROBES16-0023, 17-APRA17-0066, NNX17AJ82G, NNX13AH54G, 80NSSC18K0246, 

80NSSC18K0473, 80NSSC19K0626,
80NSSC18K0464, 80NSSC22K1488, 80NSSC19K0627 and 80NSSC22K0426, by the French space agency
CNES, and by National Science Centre in Poland grant n. 2017/27/B/ST9/02162 and 2020/37/B/ST9/01821. This research used resources of the National Energy Research Scientific Computing Center (NERSC), a U.S. Department of Energy
Office of Science User Facility operated under Contract No. DE-AC02-05CH11231. We acknowledge the
ASI-INFN agreement n. 2021-8-HH.0,  its amendments and agreement n. 2020-26-Hh.0. We acknowledge the NASA Balloon Program
Office and the Columbia Scientific Balloon Facility and staff for extensive support. 
 This research has also been
supported by the Interdisciplinary Scientific and Educational School of Moscow
University ``Fundamental and Applied Space Research'' and by Russian State
Space Corporation Roscosmos. The article has been prepared based on research
materials collected in the space experiment ``UV atmosphere''.  
 
\providecommand{\href}[2]{#2}\begingroup\raggedright\endgroup
\input{JEM-EUSO_Authors_July2023.tex}



%
%
%

\end{document}

%% file: JEM-EUSO_Authors_July2023.tex
\newpage
{\Large\bf Full Authors list: The JEM-EUSO Collaboration\\}

\begin{sloppypar}
{\small \noindent
S.~Abe$^{ff}$, 
J.H.~Adams Jr.$^{ld}$, 
D.~Allard$^{cb}$,
P.~Alldredge$^{ld}$,
R.~Aloisio$^{ep}$,
L.~Anchordoqui$^{le}$,
A.~Anzalone$^{ed,eh}$, 
E.~Arnone$^{ek,el}$,
M.~Bagheri$^{lh}$,
B.~Baret$^{cb}$
% ...
}
\end{sloppypar}
\vspace*{.3cm}

{ \footnotesize
\noindent
% ...
$^{cb}$ Universit\'e de Paris, CNRS, AstroParticule et Cosmologie, F-75013 Paris, France\\
% ...
$^{ed}$ Istituto Nazionale di Fisica Nucleare - Sezione di Catania, Italy\\
% ...
$^{eh}$ INAF - Istituto di Astrofisica Spaziale e Fisica Cosmica di Palermo, Italy\\
% ...
$^{ek}$ Istituto Nazionale di Fisica Nucleare - Sezione di Torino, Italy\\
$^{el}$ Dipartimento di Fisica, Universit\`a di Torino, Italy\\
% ...
$^{ep}$ Gran Sasso Science Institute, L'Aquila, Italy\\
% ...
$^{ff}$ Nihon University Chiyoda, Tokyo, Japan\\ 
% ...
$^{ld}$ University of Alabama in Huntsville, Huntsville, AL, USA\\
$^{le}$ Lehman College, City University of New York (CUNY), NY, USA\\
% ...
$^{lh}$ Georgia Institute of Technology, USA\\
% ...
}